# Photonic integration of an optical atomic clock

Z. L. Newman[1*], V. Maurice[1], T. E. Drake[1], J. R. Stone[1,2], T. C. Briles[1], D. T. Spencer[1], C. Fredrick[1,2], Q. Li[3], D. Westly[3], B. R. Ilic[3], B. Shen[4], M.-G. Suh[4], K. Y. Yang[4], C. Johnson[5], D. M. S. Johnson[5], L. Hollberg[6], K. Vahala[4], K. Srinivasan[3], S. A. Diddams[1,2], J. Kitching[1], S. B. Papp[1,2], M. T Hummon[1]

[1] *National Institute of Standards and Technology, Boulder, CO, 80305 USA*
[2] *Department of Physics, University of Colorado Boulder, Boulder, CO 80309, USA*
[3] *Center for Nanoscale Science and Technology, National Institute of Standards and Technology, Gaithersburg, MD, 20899 USA*
[4] *California Institute of Technology, Pasadena, CA, 91125 USA*
[5] *Charles Stark Draper Laboratories, Cambridge, MA, 02139*
[6] *Stanford University, Stanford, CA 94305*

**Abstract**: Laboratory optical atomic clocks achieve remarkable accuracy (now counted to 18 digits or more), opening possibilities to explore fundamental physics and enable new measurements. However, their size and use of bulk components prevent them from being more widely adopted in applications that require precision timing. By leveraging silicon-chip photonics for integration and to reduce component size and complexity, we demonstrate a compact optical-clock architecture. Here a semiconductor laser is stabilized to an optical transition in a microfabricated rubidium vapor cell, and a pair of interlocked Kerr-microresonator frequency combs provide fully coherent optical division of the clock laser to generate an electronic 22 GHz clock signal with a fractional frequency instability of one part in $10^{13}$. These results demonstrate key concepts of how to use silicon-chip devices in future portable and ultraprecise optical clocks.

**Main Text:**

Optical atomic clocks, which rely on high-frequency, narrow-linewidth optical transitions to stabilize a clock laser, outperform their microwave counterparts by several orders of magnitude due to their inherently large quality factors (*1*). Optical clocks based on laser-cooled and lattice-trapped atoms have demonstrated fractional instabilities at the $10^{-18}$ level (*2*), setting stringent new limits on tests of fundamental physics (*3, 4*) and may eventually replace microwave clocks in global timekeeping, navigation and the definition of the SI second (*5*). Despite their excellent performance, optical clocks are almost exclusively operated by metrological institutions and universities due to their large size and complexity.

Although significant progress has been made in reducing the size of laser-cooled atomic clocks to fit inside a mobile trailer (*6*), applications of these clocks are still limited to metrological clock comparisons and precision geodesy (*7*). In contrast, optical oscillators referenced to thermal atomic or molecular vapors can be realized in small form factors and still reach instabilities below $10^{-14}$ (*8, 9*). A fully integrated optical clock would benefit many of the applications (*10*) that currently utilize compact or chip-scale (*11*) microwave atomic clocks but, until recently, techniques for on-chip laser stabilization to atoms (*12*) and optical frequency division (*13*) were not available. Here, we propose and demonstrate an architecture for an integrated optical clock, based on an atomic vapor cell implemented on a silicon chip and a microresonator frequency comb ("microcomb") system for optical frequency division. Experimentally, this consists of a semiconductor laser local oscillator locked to the rubidium-87 two-photon transition at 385.284 THz that is coherently divided down to a 22 GHz clock tone by stabilizing a pair of interlocked microcombs to the local oscillator.

---

[*] zachary.newman@nist.gov

Microcombs, optical-frequency combs that utilize four-wave mixing of a continuous-wave pump laser inside a high-Q optical microresonator, can be made to operate over a wide range of repetition rates, from ≈10 GHz to 1 THz (*14*). As part of prior work in our group, Papp et. al. (*13*) demonstrated the first optical clock based on a microcomb by stabilizing two modes of a 33 GHz silica ($SiO_2$) microresonator to the D2 and D1 optical transitions in rubidium at 780 nm and 795 nm, respectively. However, due to the lack of self-referencing (*15*), the full stability of the optical standard was not transferred to the microwave domain. Recent progress in the field of microcombs has led to the realization of mode-locked, low-noise comb generation via temporal soliton formation in the resonator (*16–18*). These dissipative Kerr-soliton (DKS) frequency combs have been used to demonstrate octave-spanning comb operation (*19, 20*), self-referencing (*21–23*), and carrier-offset-frequency ($f_{ceo}$) phase stabilization (*24–26*). To date, octave-spanning operation has only been demonstrated using combs with ≈1 THz repetition rates, well outside the bandwidth of traditional electronic detectors, a requirement for producing a usable microwave clock signal. To circumvent this issue, our optical clock architecture employs two interlocked microcombs: a high repetition rate, octave-spanning comb used for self-referencing and a narrow-band comb used to produce an electronically detectable microwave output.

Fig. 1a shows a schematic of the experiment. The local oscillator ("clock laser") for our clock is a 778.1 nm, distributed Bragg reflector (DBR) laser that is referenced to the two-photon transition in rubidium-87 in a microfabricated vapor cell. We generate a 1 THz repetition rate, octave-spanning, DKS frequency comb by coupling ≈100 mW of pump light from a 1.54 μm external cavity diode laser (ECDL) into a $Si_3N_4$ (SiN) microresonator, which is used for coarse optical division. Full stabilization of the SiN comb is accomplished by stabilizing $f_{ceo}$ and locking a second tooth of the comb to the clock laser. We independently generate a narrow-band, 22 GHz repetition rate, DKS comb by coupling ≈160 mW of light from a 1.556 μm ECDL via tapered optical fiber into a silica microresonator. The two combs are then interlocked, and we use the silica comb as a finely spaced ruler to measure the repetition rate of the SiN comb. The output of the clock is a 22 GHz optical pulse train (and corresponding electrical signal) that is phase stabilized to the rubidium two-photon transition. The techniques for locking the DBR laser to the rubidium atoms and stabilizing the frequency combs are detailed in the Methods. Fig. 1 b, c, and d show images of the main components of the clock: the two microresonators and the Rb cell. All three elements are microfabricated devices and, in future implementations of the concept introduced here, would support more advanced integration.

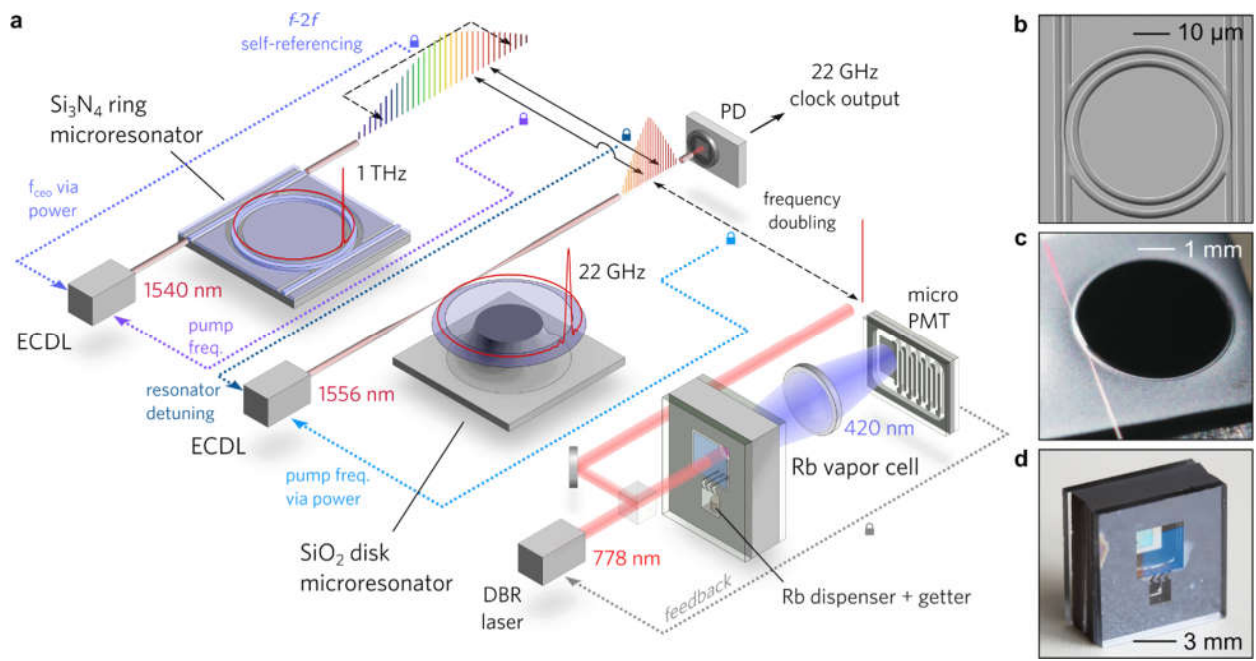

**Fig. 1. Schematic of photonic optical atomic clock.**

**a**, The microfabricated optical clock consists of an optical local oscillator, a microfabricated rubidium vapor cell, and a pair of microresonator frequency combs, which serve as optical clockwork. Absorption of the clock laser in the cell is detected via the collection of 420 nm fluorescence using a microfabricated photomultiplier tube (PMT). The optical clockwork consists of interlocked DKS combs generated using a ≈2 mm diameter, silica microresonator and a 46 μm diameter, SiN microresonator. Stabilization of the frequency combs' output is performed via electronic feedback (indicated by dotted lines) to the pump frequency and resonator detuning of the ECDLs used to pump the microresonators. The feedback signals are generated from optical heterodyne beat notes of adjacent comb teeth as indicated by the solid black arrows. In some cases, frequency doubling (dashed black arrows) was required to compare optical signals. For simplicity, we do not picture the frequency and intensity modulators used for feedback in the comb frequency servo loops. **b**, scanning-electron microscope (SEM) image of the SiN microresonator. Photographs of **c**, the silica microresonator and **d**, the microfabricated Rb vapor cell.

     The two-photon transition in rubidium (Fig. 2 inset) has been studied extensively for use as an optical frequency standard (*27–30*). Here, we only discuss details of this system relevant to spectroscopy in a microfabricated vapor cell. The clock laser is locked to the $5S_{1/2}$ (F=2) to $5D_{5/2}$ (F=4) two-photon transition in rubidium-87 at 778.106 nm (385.284566 THz) using a 3×3×3 mm vapor cell (Fig. 1d). The rear window of the cell is covered with a high-reflectivity coating (R=99.8%) which is used to retroreflect the clock laser and provide the counter-propagating beams required to excite the Doppler-free, two-photon transition. The front window is anti-reflection coated on both sides to prevent parasitic reflections.

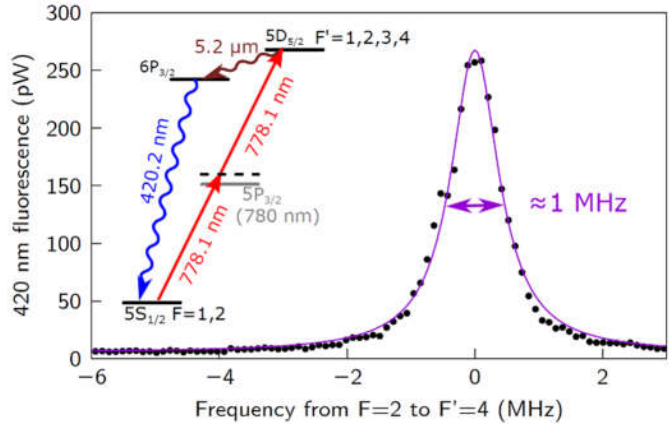

**Fig. 2. Spectroscopy of optical clock transition.**

Doppler-free fluorescence spectroscopy of the optical clock transition between the $5S_{1/2}$, F=2 to $5D_{5/2}$, F=4 levels at 385.284566 THz, with a full width at half maximum of ≈1 MHz. The atomic level structure of the rubidium-87 two-photon transition is shown as an inset.

    Excitation of the two-photon transition is detected via fluorescence at 420 nm from the 5D to 6P to 5S decay path by a microfabricated PMT along the optical axis of the clock laser. Fig. 2 shows the output of the PMT as the clock laser is swept across the clock transition. A Lorentzian fit (purple) to the fluorescence signal gives a linewidth of ≈1 MHz, which includes contributions of 330 kHz from the natural linewidth, ≈475 kHz from the laser linewidth, ≈100 kHz of transit time broadening (*31*), and ≈125 kHz due to collisional broadening from background gases in the cell (see Methods).

    Fig. 3a and 3b show optical spectra of the free-running soliton combs. During clock operation, the silica comb pump laser, $\nu_{GHz,pump}$, is frequency doubled and subsequently phase locked to the clock laser, $\nu_{Rb}$. The SiN comb tooth at 1556 nm ($\nu_{THz,pump-2}$) is then locked to $\nu_{GHz,pump}$. Here, $\nu_{GHz/THz,pump-n}$, describes the silica/SiN (GHz/THz) comb tooth *n* modes on the low frequency side of the pump. We lock the terahertz comb offset frequency, $f_{ceo}$, with an ECDL assisted, *f*-2*f* interferometer (*25*). Simultaneous stabilization of the SiN comb offset frequency and repetition rate (via $\nu_{THz,pump-2}$ locked to $\nu_{Rb}$) effectively divides the clock laser from 385 THz to 1 THz. We complete the optical frequency division by phase locking $\nu_{GHz,pump-48}$ to $\nu_{THz,pump-3}$ at 1564 nm which generates a stable, Rb-referenced, 22 GHz clock output tone. Fig. 3c shows the beat note between the 22 GHz clock output and a 22 GHz signal referenced to a hydrogen maser.

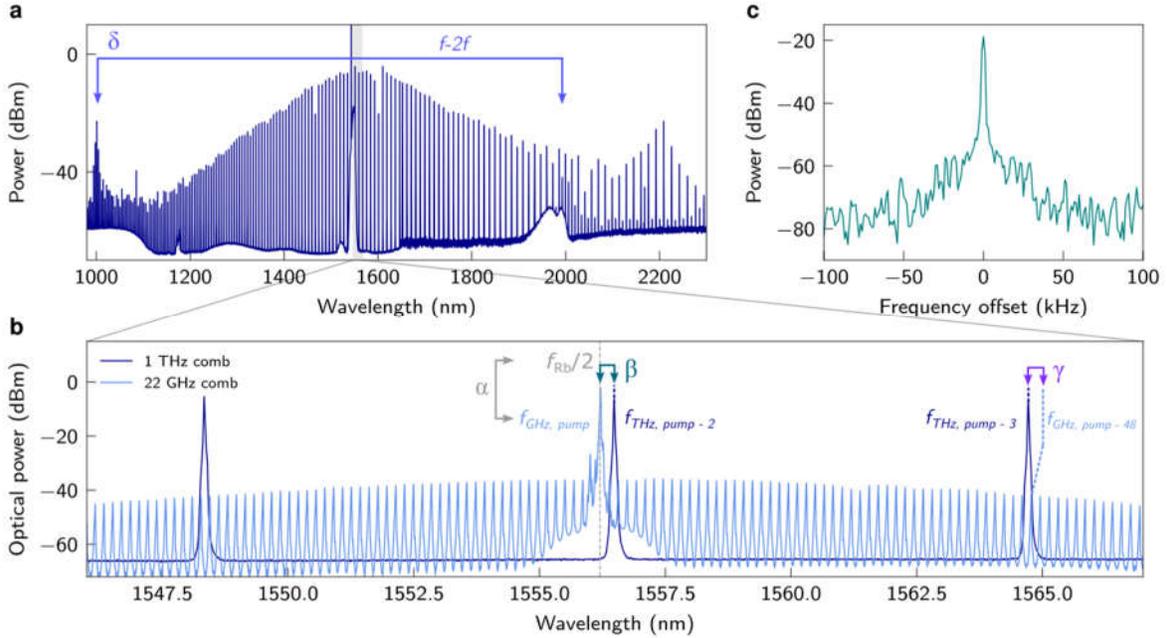

**Fig. 3. Microcomb spectra.**

**a**, SiN microresonator comb spectrum, showing the 1 THz comb tooth spacing. An *f-2f* interferometer (light blue arrow) in which light at 1998 nm is frequency doubled to heterodyne with the comb teeth in the dispersive wave at 999 nm is used to measure and stabilize the offset frequency, $f_{ceo}$, of the SiN comb ($\delta=f_{ceo}/10$ MHz). **b**, silica microresonator comb spectrum (light blue) with 22 GHz repetition rate, shown overlaid with 3 teeth from the 1 THz, SiN comb (dark blue). Arrows indicate phase locks used to stabilize the combs. Greek letters show the ratio-of-integer values multiplied by a 10 MHz clock that are used as a reference for each of the phase locks. For the devices used in the experiment $\alpha$=154.224, $\beta$=3525.29238, $\gamma$=539.9808, and $\delta$=-1280.0. **c**, RF spectrum of the 22 GHz clock output.

The frequency of the clock output, $f_{GHz,rep}$, is given by:

$$f_{GHz,rep} = \frac{\nu_{Rb} + \left(\alpha/2 + 2\cdot\left[(q-1)\cdot\beta + q\cdot\gamma - \delta\right]\right)\cdot f_{10MHz}}{2\cdot q\cdot p} \quad (1)$$

where $\nu_{Rb}$ is the clock laser frequency, $q$=190 is the mode number of the SiN comb line closest to the clock transition $\nu_{THz,pump-2}$ and $p$=48 is the number of 22 GHz comb modes separating the lock points to the SiN comb teeth, $\nu_{THz,pump-2}$ and $\nu_{THz,pump-3}$. The synthesizer frequencies used in each of the phase locks between comb teeth are defined in terms of the ratios of integers $\alpha$, $\beta$, $\gamma$, and $\delta$ relative to a 10 MHz clock referenced to a hydrogen maser, $f_{10 MHz}$, as indicated in Fig. 3B. The denominator in Eq. 1 gives the full division factor of the comb: $n = 2\times48\times190 = 18240$. A crystal oscillator referenced to the free-running repetition rate of the silica comb (Fig. 4B, open green squares) can instead be used as a reference for phase locking the microcombs and the clock can be run with no external frequency references.

Fig. 4 summarizes the performance of the optical clock. Fig. 4a shows a plot of the 22 GHz clock output measured against the hydrogen maser and multiplied up to the optical domain (blue) along with a plot of the clock laser frequency measured against an auxiliary erbium fiber frequency comb (orange) as an independent confirmation of the clock accuracy. The absolute frequency shift of our clock is $\Delta\nu \approx$-22.7 kHz ($\Delta\nu/\Delta\nu_{Rb} \approx 5\times10^{-11}$), which is primarily due to the light shift and the collision shift (see Methods). The mean values of the two measurements of the clock frequency agree to within their standard error.

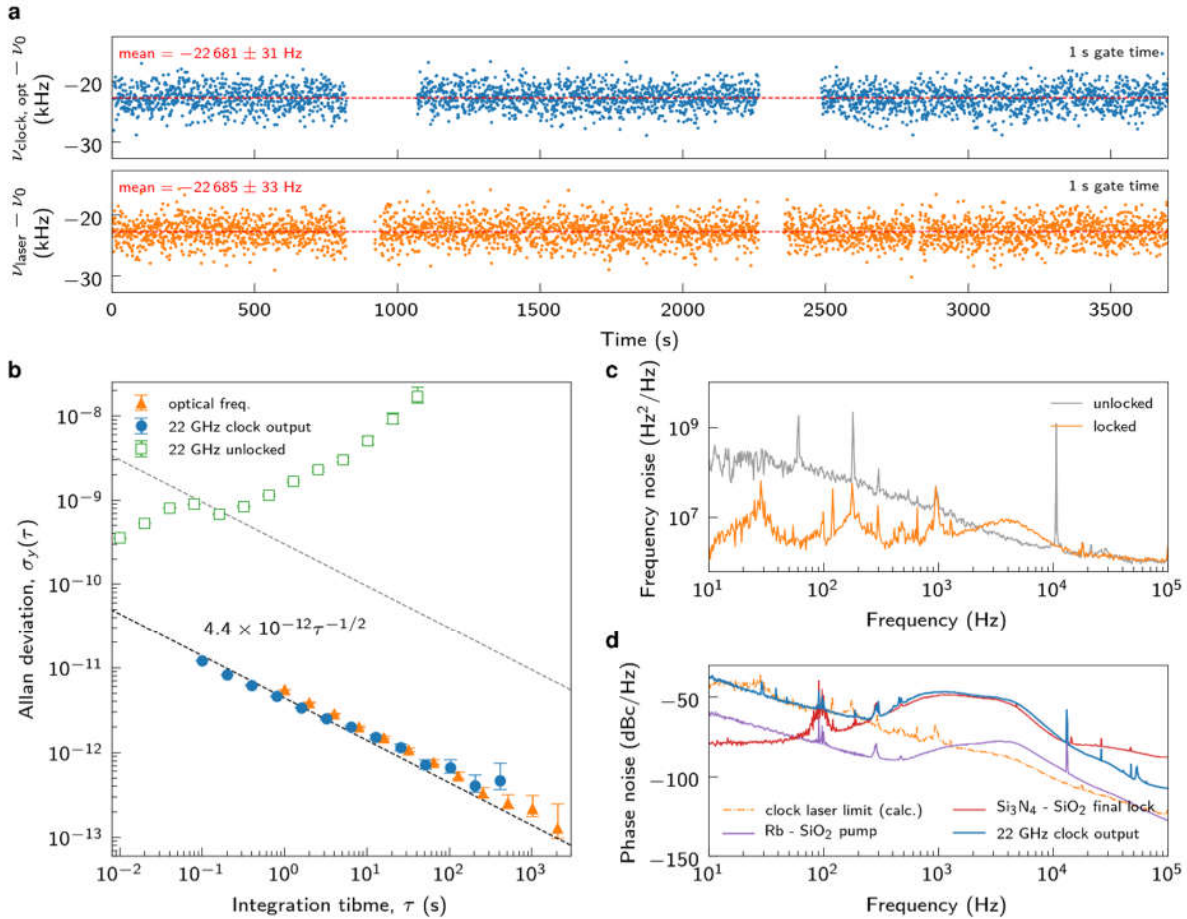

**Fig. 4. Optical clock performance.**

**a,** Time-series measurement of the clock optical frequency offset: derived from the 22 GHz clock output frequency, $f_{clock,opt}$ (top, blue) and derived from the beat note against the Er:fiber frequency comb, $f_{laser}$ (bottom, orange). The reference frequency is $\nu_0$=385284566370400 Hz (*30*). Short breaks in the data indicate periods where the clock laser dropped out of lock. **b,** Comparison of Allan deviation of 22 GHz clock output for the free-running silica microresonator (green open squares) and the fully stabilized comb (blue circles). The Allan deviation of the heterodyne beat note of the clock laser and the Er:fiber comb (orange triangles) is shown for comparison. Error bars represent a 68% confidence interval. The dotted gray shows the typical performance specifications of the Microsemi CSAC. **c,** Frequency noise spectra for the free-running and locked clock laser. A large servo bump at ≈4 kHz is evident in the phase and frequency noise spectra. **d,** Phase noise away from carrier for the 22 GHz clock output signal (blue) along with the contributions from the intermediate phase locks (red and purple) and the phase noise of the clock laser calculated from the laser frequency noise spectrum (orange dashed) which gives a lower limit on the phase noise of the 22 GHz clock output.

Fig. 4b shows an Allan deviation of the fully stabilized clock signal (Fig. 4a, blue) measured against the hydrogen maser along with an Allan deviation of the heterodyne of the clock laser with the fiber frequency comb (Fig. 4a, orange). The measured fractional frequency instability of the microwave output is $4.4\times10^{-12}/\tau^{1/2}$ and is limited by frequency noise on the DBR laser (Fig. 4b) via the intermodulation effect (*32*). At long integration times (>$10^3$ s) we notice that the drift in the clock laser frequency is correlated with the laboratory temperature and limits the ultimate stability of the clock to $\approx 1\times10^{-13}$.

Fig. 4b shows the clock phase noise (along with that of the intermediate phase locks) which is competitive with high frequency (tens of gigahertz) analog signal generators. At low frequencies (<300 Hz), the phase noise is limited by the intrinsic phase noise of the clock laser (consistent with perfect optical division at integration time >1 s), while the bump around 1 kHz results from the final phase lock between the silica and SiN combs. The phase noise at high frequencies (>10 kHz, outside the bandwidth of the comb-to-comb phase locks) is likely due to the silica comb pump laser (*33*). This suggests the clock could be operated as a low phase noise oscillator by utilizing narrow-linewidth pump sources.

In summary, we have demonstrated a compact, optical atomic clock with a stability of $4.4\times10^{-12}/\tau^{1/2}$ that makes use of a pair of self-referenced, interlocked microresonator frequency combs to transfer the stability of the optical standard down to the microwave domain. At present, the stability of our clock is limited by the performance of available, integrated 778 nm sources but short-term stabilities near $10^{-13}$ at 1 s may be possible with low-noise lasers. We anticipate that devices such as integrated narrow-linewidth ECDL with fast frequency tuning rates and waveguide-based second-harmonic generators (*34*) will eventually replace the off-the-shelf components in our clock, making a fully integrated optical atomic clock viable. State-of-the-art chip-scale optical clocks would impact applications including gravitational and remote sensing, timing and navigation when GPS is unreliable, and synchronizing large aperture networks and enable in-situ, SI-traceable calibrations of laboratory instruments.

**Methods:**

## 1. Microfabricated vapor cell

The rubidium vapor cell is composed of a 10×10×3 mm silicon frame sandwiched between two 700 μm-thick aluminosilicate glass pieces. The silicon frame is fabricated by deep reactive ion etching of a blank silicon wafer and features a main chamber 3×3 mm and an ancillary chamber 1.5×1.5 mm that are connected through 125 μm-wide baffles. The glass windows are anodically bonded to the silicon frame. Before bonding the second window under vacuum, a rubidium dispensing pill is introduced in the ancillary chamber along with a piece of non-evaporable getter. Rubidium is then released into the cell by heating the dispensing pill with a focused laser beam (*35*). Helium diffusion through the glass windows as the cell ages accounts for ≈75 kHz of broadening and ≈25 kHz of broadening is due to unwanted gases that arise during the cell bonding and filling process.

We have measured the absolute frequency of our clock to be $\nu \approx 385284566347315 \pm 30$ Hz, which corresponds to a frequency shift from the accepted value of the two-photon transition frequency of $\Delta\nu \approx -22.7$ kHz (*30*) and is primarily due to the light shift and the collision shift. We have measured the light shift to be ≈-1.5 kHz/mW resulting in a ≈-23.4 kHz shift for the 15.6 mW of clock laser power. We expect a ≈+6 kHz collision shift from helium diffusion and attribute the remaining ≈-5 kHz to collision shifts from background gases. Both the helium collision shift and the background gas collision shift are consistent with our ≈100 kHz collision broadening measurement (*29*).

## 2. Stabilized 778 nm laser system

The output of the clock laser is sent through an optical isolator and coupled into a polarization maintaining single mode optical fiber for spatial mode filtering. Following the fiber, a liquid crystal modulator is used to stabilize the optical power used to probe the rubidium atoms. The linearly polarized probe light is then focused to a spot size of 224 μm on the back window of the cell and retroreflected by the high-reflectivity coating on the back window. We typically operate the rubidium standard with ≈16 mW of optical probe power. Fluorescence at 420 nm from the two-photon transition is collected along the probe beam axis with a pair of aspheric condenser lenses and 420 nm bandpass interference filter and detected using a commercially available, microfabricated PMT.

To achieve a sufficient atomic density, the cell is heated to 100°C using a resistive heater driven with an alternating current to avoid magnetic bias fields near the cell. We stabilize the cell temperature to within 10 mK by feeding back to the amplitude of the alternating current. In addition, the heated cell mount, the PMT and the detection optics are housed inside a two-layer magnetic shield to minimize first order Zeeman shifts from stray magnetic field in the laboratory environment.

We stabilize the clock laser to the two-photon fluorescence spectrum using frequency modulation spectroscopy. The frequency of the clock laser is modulated at 10 kHz by directly modulating the DBR laser current with a modulation depth corresponding to a frequency excursion of ≈1 MHz. We generate an error signal using a lock-in amplifier to demodulate the observed fluorescence signal and lock the laser frequency by feeding back directly to the laser current.

As mentioned in the main text, we believe the short-term stability of our clock is limited by intermodulation noise from the DBR clock laser. The stability limit of a frequency standard due to the intermodulation effect can be approximated by (*32*):

$$\sigma_y \sim \left[ S_y(f_{\text{mod}})/4\tau \right]^{1/2}$$

where S($f_{\text{mod}}$) is the oscillator frequency noise at twice the modulation frequency, in this case $f_{\text{mod}}$ = 10 kHz. Fig. 4c shows a frequency noise spectrum of the DBR laser measured by beating against the fiber comb. The DBR laser frequency noise at $2f_{\text{mod}}$ is $2 \times 10^6$ Hz$^2$/Hz, which corresponds to a intermodulation limited stability of $\sigma_y \sim 2 \times 10^{-12}/\tau^{1/2}$, and is consistent with our measured short-term stability. The short-term stability can be improved by utilizing low-noise sources for the clock laser. In fact, we have measured fractional stabilities of $\approx 4.5 \times 10^{-13}/\tau^{1/2}$ using a commercial, 778 nm ECDL.

Roughly $\approx$1-2 mW of the stabilized clock laser is sent though an optical fiber and used to stabilize the pump frequency of the two microcombs. Additionally, $\approx$1 mW of the clock light is directed into a second optical fiber and beat against an auxiliary 250 MHz repetition rate, erbium fiber frequency comb to directly monitor the clock laser optical frequency.

### 3. Silica microresonator

Fig. 1c shows an image of the wedged, 21.97 GHz free-spectral range silica (SiO$_2$) microresonator used to measure the repetition rate of our self-referenced Si$_3$N$_4$ microcomb. The design, fabrication and implementation of the silica comb has been described in detail elsewhere (*36*). Briefly, the microresonator is fabricated by thermally growing a layer of SiO$_2$ on a Si substrate. The silica layer is then shaped into a wedge-resonator by lithographically patterning a photoresist material and etching the silica with a hydrofluoric acid solution. As a final step, the pedestal is formed by applying a XeF$_2$ dry etch to the Si substrate.

### 4. Si$_3$N$_4$ microresonator

Fig. 1b shows an SEM image of the SiN microresonator which is described in detail in Briles *et al.* (*25*). The Si$_3$N$_4$ resonator is fabricated by first depositing a thin layer ($\approx$615 nm) of Si$_3$N$_4$ on a thermally-oxidized silicon wafer using low-pressure, chemical-vapor deposition. The 46 μm diameter ring resonator and coupling waveguides are then patterned using electron-beam lithography and formed using a dry etching process.

As is the case with the silica resonator, the SiN comb characteristics depend strongly on the resonator dispersion, which we control via the resonator geometry. For example, the carrier offset frequency can be controlled by varying the ring radius while the spectral location of the dispersive waves (as in Fig. 2) can be controlled by varying the thickness of the Si$_3$N$_4$ layer, the width of the ring and the final resonator cladding (air or oxide layer).

### 5. Dual comb stabilization procedure

Our technique for generating single soliton combs in microresonators is described in detail in Stone *et al.* (*33*). Briefly, we create a soliton frequency comb in the resonator by sweeping the pump frequency through the cavity resonance (from blue to red) with a single side band-suppressed carrier (SSB-SC) electro-optic phase modulator. The cavity supports the formation of solitons for a limited range of pump-resonator detunings, known as the soliton existence range, which is proportional to the linewidth of the cavity (*37*, *38*). As light is coupled into the resonator, the cavity resonance frequency shifts due to the thermo-optic effect (*3*). As long as the thermal frequency drift stays within the soliton existence range, open loop control of the cavity detuning is sufficient to maintain a soliton. This is the case for the SiN resonator.

The linewidth of the silica resonator is substantially narrower, and, as a result, we lock the cavity detuning to compensate for the thermally induced shift. To accomplish this, RF sidebands are applied to the pump frequency and used to generate a Pound-Drever-Hall (PDH) error signal that is measured on a photo-detector at the output of the comb. We stabilize the soliton comb by locking the pump frequency detuning from the cavity resonance by feeding back to the setpoint of the SSB-SC modulator. The lock point of the PDH servo corresponds to the high-frequency PDH sideband and results in a pump frequency that is red-detuned from the cavity resonance by RF modulation frequency, which is required for soliton generation. The modulation frequency (and resulting cavity detuning) is chosen such that the pump frequency is nominally independent of changes in the cavity detuning. The center frequency of the soliton comb can then be controlled by thermally tuning the optical path length of the cavity with an external heater.

During clock operation we frequency double the silica comb pump light to 778 nm and beat it against the rubidium stabilized DBR laser light. The ≈1.5 GHz (*α*) beat note between the two lasers is used to phase lock the comb pump light by feeding back to an acousto-optical modulator (AOM) and controlling the intracavity power which, in turn, thermally shifts the cavity resonance. Next, we phase lock the SiN comb tooth $\nu_{THz, pump-2}$ to $\nu_{GHz,pump}$ by controlling the pump frequency with the SSB-SC (*β*). The SiN comb is self-referenced by amplifying the comb tooth at 1998 nm in a thulium-doped fiber amplifier, frequency doubling this light, beating the doubled light with the comb tooth at 999 nm and feeding back to the pump laser power with an AOM (*δ*).

Finally, we phase lock the silica comb tooth $\nu_{GHz,pump-48}$ to the $\nu_{THz, pump-3}$ tooth of the SiN comb (*γ*) by controlling the RF sideband modulation frequency, and thus, the silica comb pump detuning, which is strongly coupled to the comb repetition rate via the soliton self-frequency shift (*39*).

**Acknowledgments:**

The authors acknowledge research funding from the Defense Advanced Research Projects Agency (DARPA) Atomic Clocks with Enhanced Stability (ACES) and Direct On-Chip Digital Optical Synthesis (DODOS) programs. The views, opinions and/or findings expressed are those of the authors and should not be interpreted as representing the official views or policies of the Department of Defense or the U.S. Government. The authors would like to thank J. Burke, N. Lemke, E. Donley and T. Heavner for helpful discussions, D. Hickstein and J. McGilligan for comments on the manuscript and S. Schima, A. Dellis and D. Bopp for their help in the cell fabrication. This work is a contribution of the U.S. government (NIST) and is not subject to copyright in the United States of America. Distribution Statement 'A' (Approved for Public Release, Distribution Unlimited).